\documentclass[aps,twocolumn]{revtex4}                      
\usepackage{amsmath,amssymb}
\begin{document}
\title{Thermodynamics of Fractal Universe}
\author{Ahmad Sheykhi$^{1,2}$\footnote{sheykhi@uk.ac.ir}, Zeinab Teimoori $^{3}$ and Bin Wang $^{4}$\footnote{wang-b@sjtu.edu.cn}}
\address{$^1$ Physics Department and Biruni Observatory, College of Sciences, Shiraz University, Shiraz 71454, Iran\\\\
          $^2$  Research Institute for Astronomy and Astrophysics of Maragha
         (RIAAM), P.O. Box 55134-441, Maragha, Iran\\
          $^3$ Department of Physics, Shahid Bahonar University, P.O. Box 76175, Kerman, Iran\\
          $^4$ INPAC and Department of Physics, Shanghai Jiao Tong University, Shanghai 200240, China}

 \begin{abstract}
We investigate the thermodynamical properties of the apparent
horizon in a fractal universe. We find that one can always rewrite
the Friedmann equation of the fractal universe in the form of the
entropy balance relation $ \delta Q=T_h d{S_h}$, where $ \delta Q
$ and $ T_{h} $ are the energy flux and Unruh temperature seen by
an accelerated observer just inside the apparent horizon. We find
that the entropy $S_h$ consists two terms, the first one which
obeys the usual area law and the second part which is the entropy
production term due to nonequilibrium thermodynamics of fractal
universe. This shows that in a fractal universe, a treatment with
nonequilibrium thermodynamics of spacetime may be needed. We also
study the generalized second law of thermodynamics in the
framework of fractal universe. When the temperature of the
apparent horizon and the matter fields inside the horizon are
equal, i.e. $T=T_h$, the generalized second law of thermodynamics
can be fulfilled provided the deceleration and the equation of
state parameters ranges either as  $-1 \leq q < 0 $, $- 1 \leq w <
- 1/3$ or as $q<-1$, $w<-1$ which are consistent with recent
observations. We also find that for $T_h=bT$, with $b<1$, the GSL
of thermodynamics can be secured in a fractal universe by suitably
choosing the fractal parameter $\beta$.
\end{abstract}
\maketitle

\newpage
\section{Introduction\label{Intro}}
Nowadays, it is a general belief that there is a deep connection
between thermodynamics and gravity. The story started with the
discovery of black holes thermodynamics in $1970$'s by Hawking and
Bekenstein \cite{Haw1,Bek,Haw,Bek1}. According to their discovery,
a black hole can be regarded as a thermodynamic system, with
temperature and entropy proportional to its surface gravity and
horizon area, respectively. After that, people were speculating
that maybe there is a direct connection between thermodynamics and
Einstein equation, a hyperbolic second order partial differential
equation for the spacetime metric. In $1995$, Jacobson \cite{Jac}
was indeed able to derive the Einstein equation
 from the
requirement that the Clausius relation $\delta Q=T\delta S$ holds
for all local acceleration horizons through each spacetime point,
where $\delta S$ is one-quarter the horizon area change in Planck
units and $\delta Q$ and $T$ are the energy flux across the
horizon and the Unruh temperature seen by an accelerating observer
just inside the horizon. Jacobson's derivation of the Einstein
field equation from thermodynamics opened a new window for
understanding the thermodynamic nature of gravity. After Jacobson,
a lot of works have been done to disclose the profound connection
between gravity and thermodynamics. It was shown that the
gravitational field equations in a wide range of theories, can be
rewritten in the form of the first law of thermodynamics and vice
versa \cite{Elin,Cai1,Akb,Pad1,Pad2,Pad3,Par,Kot,Pad4}. The
studies were also generalized to the cosmological setup, where it
was shown that the differential form of the Friedmann equation in
the Friedmann-Robertson-Walker (FRW) universe can be transformed
to the first law of thermodynamics on the apparent horizon
\cite{Cai2,Cai3,CaiKim,Fro,Dan,Abdalla,WANG,Myu,Cai4,Shey1,Shey2,Shey3}.

On the other side, the second law of black hole mechanics
expresses that the total area of the event horizon of any
collection of classical black holes can never decrease, even if
they collide and swallow each other. This is remarkably similar to
the second law of thermodynamics where the area is playing the
role of entropy. Note that the second law of black hole
thermodynamics can be violated if one take into account the
quantum effect, such as the Hawking radiation. To overcome this
difficulty, Bekenstein \cite{Bek,Bek1} introduced the so-called
total entropy $S_{\rm tot}$ which is defined as
\begin{equation}
S_{\rm tot} = S_{ h}+S_{ m},
\end{equation}
where $S_{ h}$ and $S_{ m}$ are, respectively, the black hole
entropy and the entropy of the surrounding matter. According to
Bekenstein's argument, in general, the total entropy should be a
non decreasing function. This statement is known as the
generalized second law (GSL) of thermodynamics,
\begin{equation}
\triangle S_{\rm tot}\geq 0.
\end{equation}
Besides, if thermodynamical interpretation of gravity near the
apparent horizon is a generic feature, one needs to verify whether
the results may hold not only for more general spacetimes but also
for the other principles of thermodynamics, especially for the GSL
of thermodynamics. The GSL of thermodynamics is a universal
principle governing the evolution of the universe. It was argued
that in the accelerating universe the GSL is valid provided the
boundary of the universe is chosen the apparent horizon
\cite{wang1,wang2,Shey4,Shey5,Shey6}.

In this paper, we would like to extend the study to the fractal
universe. Fractal cosmology was recently proposed by Calcagni
\cite{Calc1,Calc2} for a power-counting renormalizable field
theory living in a fractal spacetime. It is interesting to see
whether the Friedmann equation of a fractal universe can be
written in the form of the first law of thermodynamics. As we will
see, in a fractal universe, the Friedmann equation can be
transformed to Clausius relation, but a treatment with
nonequilibrium thermodynamics of spacetime is needed.

In the next section we review the basic equations in the framework
of fractal cosmology. In section III, we show that the Friedmann
equation of a fractal universe can be written in the form of the
fundamental relation $ \delta Q=T_h dS_h$, where $ \delta Q $ and
$ T_h $ are, respectively, the energy flux and Unruh temperature
seen by an accelerated observer just inside the apparent horizon.
In section IV, we check the validity of the GSL of thermodynamics
for a fractal cosmology. The last section is devoted to some
concluding remarks.

\section{Fractal universe}
The total action of Einstein gravity in a fractal spacetime is
given by \cite{Calc1,Calc2}
\begin{equation}\label{action}
S=S_{G}+S_{m},
\end{equation}
where the gravitational part of the action is given by
\begin{equation}
S_{G}=\frac{1}{16 \pi G}\int d\varrho(x) \sqrt{-g}(R-2\Lambda
-\omega
\partial_{\mu}\upsilon\partial^{\mu}\upsilon),
\end{equation}
and the matter part of the action is
\begin{equation}
S_{m}=\int d\varrho (x) \sqrt{-g}  \mathcal{L}_{m}.
\end{equation}
Here $g$ is the determinant of the dimensionless metric
$g_{\mu\nu} $, $\Lambda$ and  $R$ are, respectively, the
cosmological constant and Ricci scalar. $\upsilon$ is the
fractional function and $\omega $ is the fractal parameter. The
standard measure $d^{4}x$ replaced with a Lebesgue-Stieltjes
measure $d\varrho (x) $. The derivation of the Einstein equations
goes almost like in scalar-tensor models. Taking the variation of
the action (\ref{action}) with respect to the FRW metric
$g_{\mu\nu}$, one can obtain the Friedmann equations in a fractal
universe as \cite{Calc2}
\begin{equation}\label{Fr1}
H^{2}+\frac{k}{a^{2}}+
H\frac{\dot{\upsilon}}{\upsilon}-\frac{\omega}{6}\dot{\upsilon}^{2}=
\frac{8\pi G}{3}\rho +\frac{\Lambda}{3},
\end{equation}
\begin{equation}\label{Fr2}
\dot{H}+H^{2} - H
\frac{\dot{\upsilon}}{\upsilon}+\frac{\omega}{3}\dot{\upsilon}^{2}
-\frac{1}{2}\frac{\square  \upsilon}{\upsilon}= -\frac{8\pi
G}{6}(\rho +3p)+\frac{\Lambda}{3},
\end{equation}
where $H={\dot{a}}/{a}$ is the Hubble parameter, $\rho$ and $p$
are the total energy density and pressure of the ideal fluid
composing the universe, respectively. The curvature constant
$k=0,1,-1$ corresponding to a flat, closed and open universe,
respectively. The continuity equation in a fractal universe takes
the form \cite{Calc2}
\begin{equation}\label{CONTIN}
\dot{\rho}+\left(3H+\frac{\dot{\upsilon}}{\upsilon}\right) (\rho +
p)=0.
\end{equation}
It is clear that for $\upsilon =1$, the standard Friedmann
equations are recovered. We further assume that only the time
direction is fractal, while spatial slices have usual geometry.
Indeed, in the framework of fractal cosmology, classically
fractals can be timelike $[\upsilon=\upsilon(t)]$ or even
spacelike $[\upsilon=\upsilon(x)]$ (see Ref. \cite{Calc2} for
details). These two cases lead to different classical physics, but
at quantum level all configurations should be taken into account,
so there is no quantum analogue of space or timelike fractals. In
this paper we take a timelike fractal. Thus, those parameters that
depend on time change and those parts that related to $x$ remain
fixed.

Assuming a timelike fractal profile $\upsilon= t^{-\beta}$
\cite{Calc2}, where $\beta=4(1-\alpha)$
 is the fractal dimension, the Friedmann equations (\ref{Fr1}) and (\ref{Fr2}) in the absence
of the cosmological constant can be written as
\begin{equation}\label{Fr3}
H^{2}+\frac{k}{a^{2}}-\frac{\beta}{t}H-\frac{\omega
\beta^{2}}{6t^{2(\beta +1)}}=\frac{8\pi G}{3}\rho ,
\end{equation}
\begin{equation}\label{Fr4}
\dot{H}+H^{2}-\frac{\beta}{2t}H+\frac{\beta(\beta +1)}{2t^{2}}+
\frac{\omega \beta^{2}}{3 t^{2(\beta +1)}}=-\frac{8\pi G}{6}(\rho
+ 3p),
\end{equation}
while, the continuity equation (\ref{CONTIN}) takes the form
\begin{equation}\label{cont}
\dot{\rho}+\left(3H-\frac{\beta}{t}\right) (\rho + p)=0.
\end{equation}
From the definition of the fractional integral \cite{Calc2,
Hilfer}, we know that $\alpha$ ranges as $0<\alpha \leqslant 1$.
Thus for $\alpha =1$, we obtain $\beta=0$ which physically means
that the universe does not have any fractal structure and one can
recovers the well-known Friedmann equations in standard cosmology.
As one can see from  Friedmann equations (\ref{Fr3}) and
(\ref{Fr4}), we have no limit $t\rightarrow0$ for a timelike
fractal profile, since in this case the Friedmann equations
diverge unless $\beta=0$. This implies that at the early stages of
the universe, we could not have the timelike fractal structure.

In the remaining part of this paper we show that the differential
form of the Friedmann equation (\ref{Fr3}) can be written in the
form of the fundamental relation $\delta Q= T_h dS_ h$, where
$S_h$ is the entropy  associated with the apparent horizon. We
also investigate the validity of the GSL of thermodynamics for the
fractal universe surrounded by the apparent horizon.
\section{First law of thermodynamics in fractal cosmology}
For a homogenous and isotropic FRW universe the line elements can
be written
\begin{equation}
ds^2={h}_{\mu \nu}dx^{\mu}
dx^{\nu}+\tilde{r}^2(d\theta^2+\sin^2\theta d\phi^2),
\end{equation}
where $\tilde{r}=a(t)r$, $x^0=t , x^1=r$, and $h_{\mu \nu}$=diag
$(-1, a^2/(1-kr^2))$ is the two dimensional metric. The dynamical
apparent horizon, a marginally trapped surface with vanishing
expansion, is determined by the relation $h^{\mu
\nu}\partial_{\mu}\tilde {r}\partial_{\nu}\tilde {r}=0$.
Straightforward calculation gives the apparent horizon radius for
the FRW universe as \cite{Shey5}
\begin{equation}
\label{radius}
 \tilde{r}_A=\frac{1}{\sqrt{H^2+k/a^2}}.
\end{equation}
The associated temperature $T$ with the apparent horizon is given
by
\begin{equation}\label{Tem}
T_h=\frac{1}{2\pi \tilde r_A},
\end{equation}
where $A=4 \pi \tilde r_A ^2 $ is the apparent horizon area and we
have assumed the apparent horizon radius is fixed. We shall assume
the matter source in the fractal universe has a perfect fluid form
with stress-energy tensor
\begin{equation}\label{Tensor}
T_{\mu\nu}=(\rho+p)u_{\mu}u_{\nu}+pg_{\mu\nu},
\end{equation}
where $\rho$ and $p$ are the energy density and pressure,
respectively. Assuming the total energy inside the apparent
horizon is given by $E=\rho V$, where $V=\frac{4}{3}\pi \tilde
r_A^{3} $ is the volume of the 3-sphere of radius $\tilde r_A$.
Taking differential form of energy, we can obtain the energy flux
$dE=V d \rho$. Here we have assumed the volume enveloped by the
apparent horizon is fixed during the infinitesimal internal of
time $dt$. Thus $-dE$ is actually just the heat flux $\delta Q$ in
\cite{Jac} crossing the apparent horizon within an infinitesimal
internal of time $dt$, it is not the change in the matter energy
inside the apparent horizon due to the volume change, so there is
no term of volume change. Hence we can write
\begin{equation}
dE=V\dot{\rho} dt.
\end{equation}
Note that by taking $\upsilon=t^{-\beta} $, from the continuity
equation (\ref{cont}) we have $\rho\sim (a^3
\upsilon(t))^{-(1+w)}$, which shows that the measure weight
$\upsilon$ is hidden in $\rho$  and thus it is not necessary to
consider its variation in Eq. (16) separately.

Substituting $\dot{\rho}$ from the continuity equation
(\ref{cont}), we get
\begin{equation}\label{dE}
dE= -4\pi H \tilde r_A ^{3}(\rho +p) dt +\frac{4\pi}{3}
\frac{\beta}{t}(\rho +p)\tilde r_A ^{3} dt.
\end{equation}
Differentiating Friedmann equation (\ref{Fr3}) with respect to the
cosmic time $t$ and using the continuity equation(\ref{cont}), we
find
\begin{eqnarray}\label{fried2}
&&H\left(\dot{H}-\frac{k}{a^{2}}\right)-\frac{\beta}{2t}\dot{H}+\frac{\beta}{2
t^{2}}H+\frac{\omega \beta^{2}(\beta +1)}{6t^{2\beta +3}}
= \nonumber \\ && -4\pi G H (\rho +p)+\frac{4\pi
G}{3}\left(\frac{\beta}{t}\right)(\rho +p).
\end{eqnarray}
Multiplying the factor $(-\tilde r_A^{3})$ on both sides of
Eq.(\ref{fried2}), we reach
\begin{eqnarray}\label{fried3}
 -H\left(\dot{H}-\frac{k}{a^{2}}\right)\tilde
r_A^{3}+\frac{\beta}{2t}\dot{H}\tilde r_A^{3}-\frac{\beta}{2
t^{2}}H\tilde r_A^{3}
 - \frac{\omega \beta^{2}(\beta +1)}{6 t^{2\beta +3}}\tilde r_A^{3} \nonumber \\
=4\pi G H (\rho +p) r_A^{3} - \frac{4\pi G}{3}\frac{\beta}{t}(\rho +p)\tilde r_A^{3}. \nonumber
\\
\end{eqnarray}
Differentiating Eq.(\ref{radius}) with respect to the cosmic time
$t$, we obtain
\begin{equation}\label{radius1}
\dot{\tilde r}_{A}= - H\left(\dot{H}-\frac{k}{a^{2}}\right)\tilde r_A^{3}.
\end{equation}
Substituting Eq.(\ref{radius1}) into (\ref{fried3}) we can rewrite
it as
\begin{eqnarray}\label{fried4}
&&\frac{1}{G} \left [d\tilde r_A -\beta \left(\frac{H}{2t^{2}} - \frac{\dot{H}}{2t}\right)\tilde r_A^{3} dt -\frac{\omega}{6}\frac{\beta^{2}(\beta +1)}{t^{2\beta +3}}\tilde r_A^{3} dt\right] \nonumber \\ && =4\pi H
(\rho +p)\tilde r_A^{3} dt - \frac{4\pi}{3}\frac{\beta}{t}(\rho
+p)\tilde r_A^{3} dt.
\end{eqnarray}
Combining Eq. (\ref{dE}) with  (\ref{fried4}) and using the fact
that $\delta Q= -dE$ is just the energy flux crossing through the
apparent horizon, we have
\begin{equation}\label{thermo}
\delta Q= \frac{1}{G}\left[d\tilde r_A -\beta
\left(\frac{H}{2t^{2}} - \frac{\dot{H}}{2t}\right)\tilde r_A^{3}
dt - \frac{\omega}{6}\frac{\beta^{2}(\beta +1)}{t^{2\beta +3}}\tilde r_A^{3} dt\right].
\end{equation}
Eq. (\ref{thermo}) can be further rewritten as
\begin{eqnarray}\label{thermo1}
\delta Q=\frac{1}{2\pi \tilde r_A}\frac{2\pi \tilde
r_A}{G}\left[d\tilde r_A - \beta
\left(\frac{H}{2t^{2}} - \frac{\dot{H}}{2t}\right)\tilde r_A^{3}dt \right.\nonumber\\
\left. - \frac{\omega}{6}\frac{\beta^{2}(\beta +1)}{t^{2\beta +3}}\tilde r_A^{3} dt\right].
\end{eqnarray}
Using definition (\ref{Tem}) for the temperature, one can see that
Eq. (\ref{thermo1}) is just the entropy balance relation,
\begin{equation}\label{thermo2}
\delta Q=T_h dS_h,
\end{equation}
provided we define
\begin{equation}\label{entro}
d{S_h}=\frac{2\pi \tilde r_A d\tilde r_A}{G}+\frac{2\pi
\beta}{G}\left[ \left(\frac{\dot{H}}{2 t}-\frac{H}{ 2t^{2}}
\right)- \frac{\omega}{6}\frac{\beta(\beta +1)}{t^{2\beta
+3}}\right]\tilde r_A^{4} dt .
\end{equation}
Integrating, we find
\begin{eqnarray}
S_h=\frac{A}{4G}+\frac{2\pi \beta}{G} \int\left[
\left(\frac{\dot{H}}{2 t}-\frac{H}{ 2t^{2}} \right)
-\frac{\omega\beta(\beta +1)}{6t^{2\beta +3}}\right]\tilde r_A^{4}
dt.\nonumber \\
\end{eqnarray}
As one can see, in a fractal universe, the entropy $S_h$
associated with the apparent horizon consists two parts, the first
one obeys the usual area law and the second part is the entropy
term developed internally in the system as a result of being out
of equilibrium \cite{Elin,noneq}. The entropy production rate
vanishes for standard cosmology where $\beta=0$. It is worth
mentioning that even in standard cosmology one can still have
non-equilibrium thermodynamics depending on the assumptions
\cite{Elin}.
\section{GSL of thermodynamics in fractal universe\label{GSL}}
In this section we investigate the validity of the GSL of
thermodynamics in a region enclosed by the apparent horizon in the
framework of the fractal universe. Let us put $k=0$ for
simplicity, so we have ${\tilde r}_A = {1}/{H}$, $ \dot{H}=
-{\dot{\tilde r}_{A}}/{\tilde r_A^{2}}$. The total entropy
associated with the apparent horizon, $S_h$, can be written
\begin{eqnarray}\label{entro1}
d S_h= \frac{2\pi}{G}\left[\tilde r_A d\tilde r_A
-\frac{\beta}{2t}\tilde r_A^{2}d{\tilde r_A }-\frac{\beta}{2 t^2}\tilde r_A^{3}dt \right.\nonumber
\\ \left.
- \frac{\omega}{6}\frac{\beta^{2}(\beta +1)}{t^{2\beta +3}}\tilde r_A^{4} dt \right].
\end{eqnarray}
Dividing Eq.(\ref{entro1}) by $dt$, we arrive at
\begin{eqnarray}\label{entro2}
\dot{S}_h=\frac{2\pi }{G}\left[\tilde r_A \dot{\tilde r}_{A}
-\frac{\beta}{2t}\tilde r_A^{2}\dot{\tilde r}_{A}
-\frac{\beta}{2t^{2}}\tilde r_A^{3}\right.\nonumber \\
\left. -\frac{\omega}{6}\frac{\beta^{2}(\beta +1)}{t^{2\beta +3}}\tilde r_A^{4}\right].
\end{eqnarray}
Eq. (\ref{entro2}), can be written as
\begin{eqnarray}\label{etropy}
\dot{S}_h=\frac{2\pi }{G}\tilde r_A\dot{\tilde
r}_{A}\left(1-\frac{\beta}{2t}\tilde r_A \right) - \frac{2\pi
}{G}\frac{\beta}{2t^{2}} \tilde r_A^{3} \nonumber \\
 -\frac{2\pi }{G}\frac{\omega}{6}\frac{\beta^{2}(\beta +1)}{t^{2\beta +3}}\tilde r_A^{4}.
\end{eqnarray}
The Friedmann equation (\ref{Fr3}), for a flat universe, becomes
\begin{equation}
\frac{1}{\tilde r_A^{2}} - \frac{\beta}{t}\frac{1}{\tilde r_A}
-\frac{\omega \beta^{2}}{6t^{2(\beta +1)}}
= \frac{8\pi G}{3}\rho.
\end{equation}
Differentiating the above equation with respect to the cosmic time
and using the continuity equation(\ref{cont}), after some
simplification, we get
\begin{eqnarray}\label{entro3}
&&\frac{\dot{\tilde r}_{A}}{\tilde r_A^{3}}\left(1-
\frac{\beta}{2t}\tilde r_A\right)
-\frac{\beta}{2t^2}\frac{1}{\tilde r_A}
- \frac{\omega}{6}\frac{ \beta^{2}(\beta+1)}{t^{2\beta +3}}\nonumber \\ && = 4\pi G
H(\rho +p)- \frac{4\pi G}{3} \frac{\beta}{t}(\rho +p).
\end{eqnarray}
Solving this equation for $\dot{\tilde r}_{A}$, we find
\begin{eqnarray}\label{entro4}
\dot{\tilde r}_{A}=\left(1- \frac{\beta}{2t}\tilde
r_A\right)^{-1}\tilde r_A^{3}
\left[4\pi GH(\rho +p) \right.    \nonumber \\
\left. - \frac{4\pi G}{3} \frac{\beta}{t}(\rho +p)
+\frac{\beta}{2t^2}\frac{1}{\tilde r_A}
+ \frac{\omega}{6}\frac{ \beta^{2}(\beta+1)}{t^{2\beta +3}}\right].
\end{eqnarray}
Substituting $\dot{\tilde r}_{A}$ from Eq.(\ref{entro4}) into
(\ref{etropy}), we obtain
\begin{equation}
\dot{S}_h= \frac{2\pi}{G}H^{-4} \left[4\pi G H(\rho +p)-
\frac{4\pi G}{3} \frac{\beta}{t}(\rho +p)\right],
\end{equation}
which can also be rewritten in the following form
\begin{equation}
\dot{S}_h= 8\pi^2H^{-3}(\rho +p)\left(1- \frac{\beta}{3Ht}\right),
\end{equation}
where we have used ${\tilde r}_A = {H}^{-1}$ for the flat
universe. Let us discuss the two cases in which
$\dot{S_{h}}\geq0$. In the first case we assume the dominant
energy condition valid, $\rho+p\geq0$, therefore
$\dot{S_{h}}\geq0$, provided $\beta\leq3H t$. However, in an
accelerating universe the dominant energy condition may violate,
$\rho+p<0$. In this case $\dot{S_{h}}\geq0$ provided
$\beta\geq3Ht$. However, as we will see below, the GSL can be
still fulfilled in an accelerating fractal universe.

For latter convenience we also calculate $T_h \dot{S_h}$,
\begin{equation}\label{TdSh}
T_h \dot{S_h}=4\pi H^{-2}(\rho +p)\left(1-
\frac{\beta}{3Ht}\right).
\end{equation}
Next, we study the GSL of thermodynamics, namely the time
evolution of the total entropy including the entropy $S_h$
associated with the apparent horizon together with the matter
field entropy $S_{\rm in}$ inside the apparent horizon. The
entropy of the universe inside the horizon can be related to its
energy and pressure in the horizon by the Gibbs equation
\cite{Pavon2}
\begin{equation}\label{Gib2}
T dS_{\rm in}=d(\rho V)+pdV=V d\rho+(\rho+p)dV.
\end{equation}
We assume the temperature of the perfect fluid inside the apparent
horizon scales as the temperature of the horizon, which for flat
universe is  $T_{h} =H/(2\pi)$. In general, if the temperature of
the horizon differs much from that the fluid, then the energy
would spontaneously flow between the horizon and the fluid,
something at variance with FRW universe \cite{Pavon2, Muba}. Thus
we suppose that the temperature $T_h$ associated with the apparent
horizon is $T_h=bT $ \cite{wang2}, where $b$ is a real
proportional constant. Since at the present time the horizon
temperature is lower than that of the CMB by many orders of
magnitude, we will not consider the case $b > 1$.  We limit
ourselves to the assumption of the local equilibrium hypothesis,
that the energy would not spontaneously flow between the horizon
and the fluid.  Indeed, this will certainly be the situation at
late times, that is when the universe fluids and the horizon will
have interacted for a long time, it is ambiguous if it will be the
case at early or intermediate times \cite{jamil1}. However, in
order to avoid nonequilibrium thermodynamical calculations, which
would lead to lack of mathematical simplicity and generality, the
assumption of equilibrium, although restricting, has widely
accepted for studying the GSL in the literature \cite{jamil1}.
Thus, we follow this assumption and we notice that our results are
valid only at the late stages of the universe evolution where the
universe fluids and the horizon will interact for a long time.

Let us first consider the case where $b=1$. Physically, this means
that we have assumed during the infinitesimal internal time $dt$,
the temperature of the perfect fluid is equal to the temperature
$T_h$ associated with the apparent horizon. Therefore, from Gibbs
equation, after using the continuity equation (\ref{cont}), we get
\begin{eqnarray}\label{TdSm}
T \dot{S}_{\rm in}&=& -4\pi H^{-2}(\rho +p)\left(1-
\frac{\beta}{3Ht}\right) \nonumber
\\ &&+
 4\pi H^{-2}(\rho + p)\dot{\tilde r}_{A}.
\end{eqnarray}
To check the GSL of thermodynamics, we have to examine the
evolution of the total entropy $S_h + S_{\rm in}$. Adding equations
(\ref{TdSh}) and (\ref{TdSm}),  we get
\begin{equation}\label{tentropy}
T_h(\dot{S}_{h}+ \dot{S}_{\rm in})= 4\pi H ^{-2} (\rho +p)
\dot{\tilde r}_{A}= A (\rho +p)\dot{\tilde r}_{A},
\end{equation}
where $A=4\pi H^{-2}$ is the apparent horizon area. The
deceleration parameter $q$ can be written as
\begin{equation}
q=-1 - \frac{\dot H}{H^ {2}}.
\end{equation}
Using the fact that in flat FRW universe we have $\dot{\tilde
r}_{A}=-\dot{H}/H^2$, Eq.(\ref{tentropy}) can be written as
\begin{equation}
T_h(\dot{S}_{h}+ \dot{S}_{\rm in})= A \rho(1+w)(1+q),
\end{equation}
where $w={p}/{\rho} $ is the equation of state parameter. From the
above equation we see that for $(1 +q)(1+w)\geq0 $  the GSL  of
thermodynamics is fulfilled in a region enclosed by the apparent
horizon. Let us consider two cases separately.  In the first case
where $q\geq -1 $ and $w\geq -1$, the GSL holds. Recent
observations from type Ia supernova show that our universe is
currently undergoing a phase of accelerated expansion. For an
accelerating universe we have $q< 0$ and $w < - 1/3$. Therefore,
in an accelerating universe the GSL is fulfilled for $-1 \leq q <
0 $ and $- 1 \leq w < - 1/3$ which are consistent with recent
observations. In the second case where $q < -1$ and $w <-1$, the
GSL is again fulfilled, but in this case the equation of state
parameter crosses the phantom line, $w=-1$, which again some
cosmological data confirm it.
Now we consider the general case where $b<1$. Since at the present
time the horizon temperature is lower than that of the CMB by many
orders of magnitude, we will not consider the case $b > 1$. Adding
equations (\ref{TdSh}) and (\ref{TdSm}) with $T_{h}=bT$, we obtain
\begin{eqnarray}\label{tentropy1}
&&T_h (\dot{S}_{h} +\dot{S}_{\rm in})=4\pi H^{-2} (\rho
+p)\left(1-\dfrac{\beta}{3Ht}\right)(1-b) \nonumber \\ &&+4\pi b
H^{-2}(\rho +p)(1+q)  \nonumber \\
&=& 4\pi  H^{-2}(\rho
+p)\left[\left(1-\dfrac{\beta}{3Ht}\right)\left(1-b\right)
+b(1+q)\right] \nonumber \\
&=&4\pi  H^{-2}\rho(1+
w)\left[1+bq-\dfrac{\beta}{3Ht}\left(1-b\right)\right].
\end{eqnarray}
Equation (\ref{tentropy1}) shows that for $w\geq -1$, the GSL can
be secured provided that
\begin{eqnarray}\label{condition1}
\beta(1-b)\leq 3Ht(1+bq).
\end{eqnarray}
Thus, for $b< 1$, Eq. (\ref{condition1}) can be translated to
\begin{eqnarray}\label{condition2}
\beta \leq 3Ht \left(\frac{1+bq}{1-b}\right).
\end{eqnarray}
Since $\beta$  is a positive definite, that is $\beta>0$, hence
the above condition can also be written
\begin{eqnarray} \label{condbeta2}
 0<\beta\leq3Ht
\left(\frac{1+bq}{1-b}\right).
\end{eqnarray}
The right hand side of inequality (\ref{condbeta2}) should be
positive, which leads to $q>-b^{-1}$. In an accelerating universe
$q<0$, and hence the condition for keeping the GSL in a fractal
universe reduced to $-b^{-1}<q<0$.

On the other hand, if equation of state parameter crosses the
phantom line, which some observational data support it, then we
have $w<-1$. In this case, the GSL holds  provided
\begin{eqnarray}\label{condition3}
\beta(1-b)> 3Ht(1+bq).
\end{eqnarray}
 For $b<1$, the GSL can be fulfilled if
\begin{eqnarray}
\beta > 3Ht \left(\frac{1+bq}{1-b}\right).
\end{eqnarray}
We conclude that the requirement of the GSL of thermodynamics in a
fractal universe leads to constraint on the fractal dimension
parameter $\beta$. It is worth noting that our local equilibrium
hypothesis, which we used for checking the GSL, only valid at the
late time. Thus, the result obtained in this secion have no
$t\rightarrow0$ limit.

Finally, it is instructive to discuss the sign of the entropy and
temperature, when the universe lies in the phantom phase $(\omega
<-1)$. A lot of works have been done in the literature, showing
that in the absence of chemical potential in phantom regime, the
temperature must be negative, while the energy density and the
entropy  should be positive \cite{Myn,Gon}. In \cite{Lim, Pere} a
negative chemical potential was assumed for the phantom fluid, and
showed that the temperature, entropy and density must be positive.
In \cite{Sar} it was shown that with an arbitrary chemical
potential the density and entropy are always positive, while the
temperature of a phantom universe ($w<-1$) is negative, and that
of quintessence universe ($w>-1$) is positive. The temperature
negativity can only be interpreted in the quantum framework
\cite{Sar}. In \cite{Pavon2} it was found that the phantom
temperature is positive and its entropy negative. Finally, in
\cite{Bre, Noj} it was argued that one can describe the phantom
universe either with negative temperature and positive entropy, or
with negative entropy and positive temperature. Since the horizon
temperature is always positive, it is deduced that the universe
temperature will be positive even if it lies in the phantom phase.
Thus we should have a negative universe entropy in this case.
Although the total entropy is always positive. Because negative
entropy of the universe ingredients is overcome by the positive
horizon entropy.
\section{concluding remarks\label{Con}}
In summary, in this paper we studied  thermodynamics of the
apparent horizon in the framework of fractal cosmology. We found
that the Friedmann equation of a fractal universe can be
transformed to the fundamental relation $ \delta Q=T_h d{S_h}$ on
the apparent horizon, where $ \delta Q $ and $ T_h $ are the
energy flux and Unruh temperature seen by an accelerated observer
just inside the apparent horizon. We showed that the entropy $S_h$
consists two terms, the first one which obeys the usual area law
and the second part which is the entropy production term and
appears due to the nonequilibrium thermodynamics of the fractal
universe. This indicates that in a fractal universe, a treatment
with nonequilibrium thermodynamics of spacetime maybe needed. We
also investigated the time evolution of the total entropy
including the entropy $S_h$ associated with the apparent horizon
together with the matter field entropy $S_{\rm in}$ inside the
apparent horizon. We assumed the temperature of the apparent
horizon is proportional to the matter fields temperature inside
the horizon, i.e. $T_{h}=bT$. We studied several cases including
$b<1$ and $b=1$, in which the GSL of thermodynamics can be secured
in a fractal universe. Interestingly enough, we found that for an
accelerating fractal universe the GSL can be preserved, at least
for the late times where the local equilibrium hypothesis holds.
The fulfillment of the GSL of thermodynamics in an accelerating
fractal universe leads also to constraints on the fractal
dimension parameter $\beta$.
\acknowledgments{We thank the anonymous referee for constructive
comments which helped us to improve the paper significantly. This
work has been supported financially by Research Institute for
Astronomy and Astrophysics of Maragha (RIAAM) and also by Shiraz
University Research Council.}

\end{document}